\documentclass[conference]{IEEEtran}
\IEEEoverridecommandlockouts
\usepackage{cite}
\usepackage{amsmath,amssymb,amsfonts}
\usepackage{algorithmic}
\usepackage{graphicx}
\usepackage{textcomp}
\usepackage[table,xcdraw]{xcolor}
\usepackage{url}
\usepackage{multirow}
\usepackage{booktabs}
\usepackage{balance}
\usepackage{caption}
\usepackage{pgfplots}
\usepackage{balance}
\usepackage{subfig}

\begin{document}

\title{Heart Abnormality Detection from Heart Sound Signals using MFCC Feature and Dual Stream Attention Based Network}

\author{\IEEEauthorblockN{
        Nayeeb Rashid$^{1}$,
        Swapnil Saha$^{1}$,
        Mohseu Rashid Subah $^{1}$,
        Rizwan Ahmed Robin $^{1}$, \\
        Syed Mortuza Hasan Fahim $^{1}$,
        Shahed Ahmed$^{1}$
        and Talha Ibn Mahmud$^{1}$}

\IEEEauthorblockA{$^{1}$Department of Electrical and Electronic Engineering (EEE)\\ 
Bangladesh University of Engineering and Technology (BUET), Dhaka - 1205, Bangladesh. 
}}%

\maketitle

\begin{abstract}
Cardiovascular diseases are one of the leading cause of death in today's world and early screening of heart condition plays a crucial role in preventing them. The heart sound signal is one of the primary indicator of heart condition and can be used to detect abnormality in the heart. The acquisition of heart sound signal is non-invasive, cost effective and requires minimum equipment. But currently the detection of heart abnormality from heart sound signal depends largely on the expertise and experience of the physician. As such an automatic detection system for heart abnormality detection from heart sound signal can be a great asset for the people living in underdeveloped areas. In this paper we propose a novel deep learning based dual stream network with attention mechanism that uses both the raw heart sound signal and the MFCC features to detect abnormality in heart condition of a patient. The deep neural network has a convolutional stream that uses the raw heart sound signal and a recurrent stream that uses the MFCC features of the signal. The features from these two streams are merged together using a novel attention network and passed through the classification network. The model is trained on the largest publicly available dataset of PCG signal and achieves an accuracy of $87.11\%$, sensitivity of $82.41\%$, specificty of $91.8\%$ and a MACC of $87.12\%$.  

\end{abstract}


\section{Introduction}

Cardiovascular disease (CVD) is one of the main reasons for mortality in the world \cite{medina2020use,perpetuini2020photoplethysmographic}. In the early stage of cardiovascular disease, heart sound auscultation is the most common primary screening tool to differentiate abnormal heart sounds from normal ones. The heart sound is created due to the opening and closing of the valves in the cardiovascular system resultant from the hemodynamics and electrical activity of the heart muscle \cite{kumar2011noise}.  Manual heart sound auscultation to detect abnormality is time-consuming, subjective, and requires extensive experience and training \cite{roy2002helping}. This makes it inevitable to develop an automatic abnormal heart sound detection system.\\

Although the work on automated classification of pathology in heart sound recordings has been performed for over 50 years, still there is some scope to be improved like extraction of new features, use the state-of-the-art artificial neural network algorithms, etc. Gerbarg et al who attempted first to classify pathology in PCGs used a threshold-based method \cite{GERBARG1963393}. Referring to the recent work, we can divide the classification approach into 2 subgroups. One group \cite{potes2016ensemble,he2021research} uses the direct raw PCG signal for classification and another group \cite{deng2020heart,li2020classification} generates features from the raw data signal and used them for classification. In the context of using raw signal data, we need to segment the cardiac cycle into four corresponding states: first heart sound (S1), systolic period (sys), the second heart sound (S2), and diastolic period (dia). These cycles are then fed to the machine learning model / deep learning model for classification. In the context of feature, these are some important signal features used in heart sound analysis: wavelet features \cite{choi2008comparison}, time, frequency, and complexity-based features \cite{clifford2012signal}, and time-frequency features \cite{son2018classification}. These features are the feed to the machine learning and ANN model for classification. Fan Li et al.\cite{li2020classification} used a 1d convolution neural network to train a total of 497 features extracted from time, frequency, and energy domain. Potes et al. \cite{potes2016ensemble} and Yaseen et al.\cite{son2018classification} used the same types of features to train SVM (support vector machine) model, Muqing Deng et al.\cite{deng2020heart} used improved MFCC features to train Recurrent Neural Network. 

\begin{figure}[t]
    \centering
    \includegraphics[width=\linewidth]{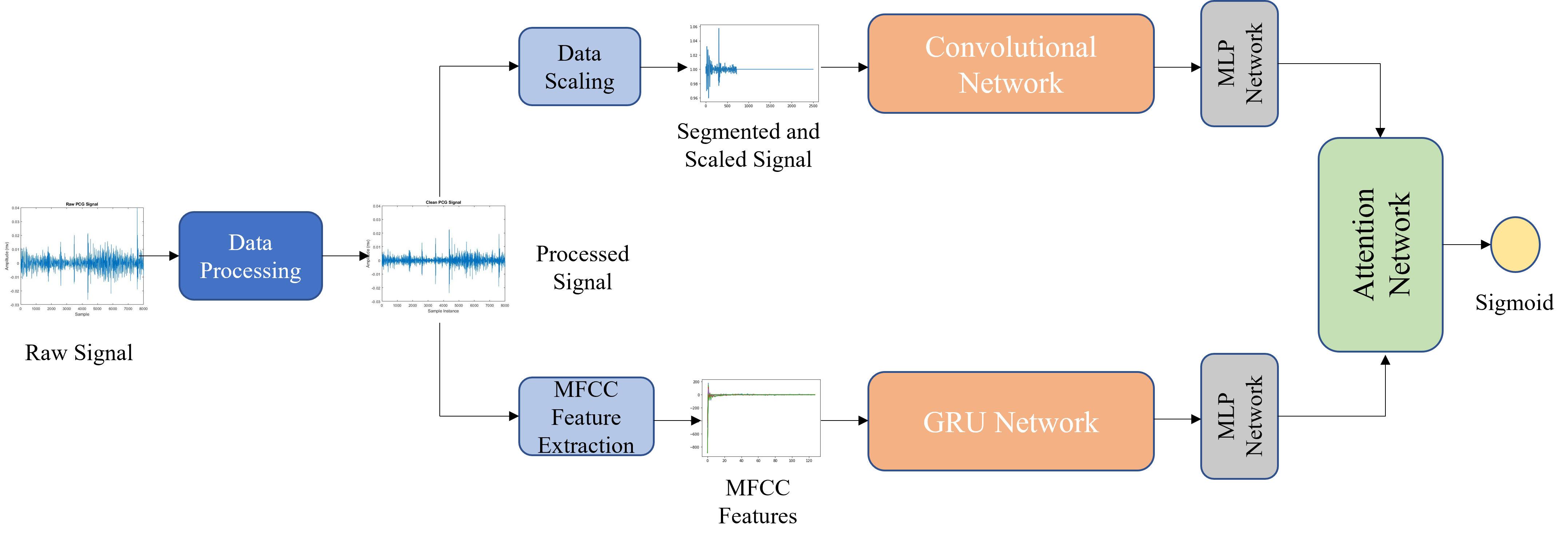}
    \caption{Model Architecture and Workflow}
    \label{flow_chart}
\end{figure}

\begin{figure*}[t]
    \centering
    \includegraphics[width=\linewidth]{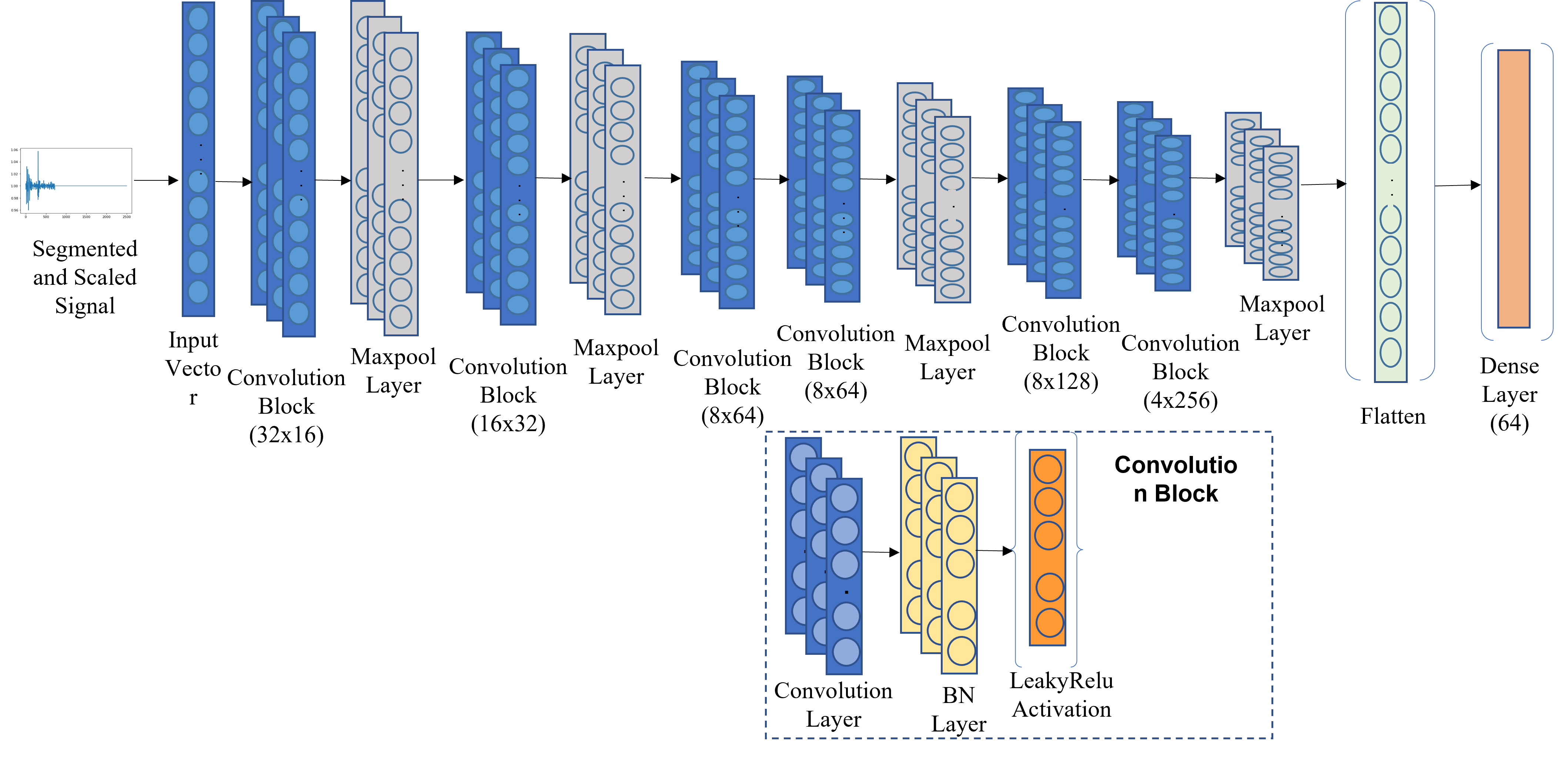}
    \caption{The detailed network architecture of the convolutional neural network}
    \label{conv}
\end{figure*}

In this project we propose a dual stream attention based deep neural network system to automatically detect abnormality in a patient’s heart sound signal. The proposed system first segments and processes the audio signal and then the  processed signal is passed to a dual stream network. One of the stream is the convolutional stream that uses the raw signal and passes it through a 1D convolutional neural network model followed by a MLP network. The other stream is the recurrent stream that first derives the MFCC features of the audio signal and passes these features to a recurrent neural network model followed by a MLP network. The features from these two streams are then combined using a novel attention network that merges these features and passes them on to the classification layer to make the final classification. The entire network is trained on a large dataset of PCG signal to perform binary classification of normal and abnormal heart sound signal.

\section{Methodology}
In the proposed method the heart sound signal is first passed through a data preprocessig block that performs various data processing on the data and prepares it for passing on the deep learning model. The processed data is then passed through two parallel network streams. The raw data is passed through the convolutional stream, while in the recurrent stream first MFCC features are extracted from the data and then it is passed to a GRU model. The features from both the streams are merged using an attention network and passed through to the classification layer. The proposed system is visually presented in Figure \ref{flow_chart}. 

\subsection{Signal Preprocessing}
The signal preprocessing block performs various data processing on the heart sound signal so that it can be passed on to the deep learning model. First in order to convert all the data in the dataset to a common sampling rate the data is downsampled to 1000 Hz. One of the major problems associated with the heart sound data is the presence of noise in the signal. So the signals are first passed through a band pass filter and spike removal filter for noise removal. Now the signals in the dataset are of variable length but the input to the model has to be of an equal length for all the data. In order to do that the signals are segmented to heart sound cycles using the methods proposed by Springer et al. \cite{springer2015logistic}. After the segmentation all the heart sound cycles are either truncated or zero padded to a fixed length of 2500 samples depending on the length of their respective cycle length. 

\subsection{Convolutional Stream}

\subsubsection{Data Scaling}
One of the issues with this PCG data is that it each cycle has zero values for almost more than half of their signal length. But convolutional neural network tends to learn better data representation when the data is non zero. So in oder to improve the performance of the convolutional network each cycle is added with an offset of one and scaled up.

\subsubsection{Convolutional Network}
In the convolutional stream we implemented a 1D convolutional neural network that is trained using the raw heart sound signal. The convolutional neural network is made up of a series of convolution blocks. Each block is composed of a 1D convolutional layer, 1D batch-normalization layer and a LeakyRelu activation function. The convolutional blocks have 1D maxpooling layers between them. The convolutional network is presented in Figure \ref{conv}.

As it can be seen from the figure, the network is designed in such a way that as it gets deeper the deeper the kernel size of the convolution block decreases while the number of filters increases. The first convolutional block has a kernel size of 32 and 16 filters. It is followed by a maxpooling layers. The pooling layers downsamples the signal to a reduced feature space. After that it is again passed through a convolutional block with kernel size of 16 and 32 filters which is followed by a pooling layer. Then there are two consecutive convolutional blocks both with a kernel size of 8 and 64 filters. After that we perform a pooling operation and again pass the signal through two consecutive convolutional blocks with kernel size of 8 and 4, and filter numbers of 128 and 256 respectively. Then we perform the final maxpool operation on the data and then the feature space is flattened and passed through a fully connected layer with 64 units.

\subsection{Recurrent Stream}

\subsubsection{MFCC Feature Extraction}
At the start of the recurrent stream the MFCC features are extracted from the raw heart sound signal data, as the network uses these MFCC features as input to the model instead of the raw signal. The Mel-frequency cepstrums or MFCC feature is generated by windowing the signal, applying DFT, taking the log of the magnitude and then warping the frequencies on a Mel scale which is followed by the inverse DCT on the data \cite{rao2017speech}. The mel scale frequecy is determined using the Equation \ref{mel}.
\begin{equation}
Mel(f) = 2595l\lg \left ( 1 + \frac{f}{700} \right )
\label{mel}
\end{equation}

\subsubsection{GRU Network}
\begin{figure}[h]
    \centering
    \includegraphics[width=0.8\linewidth]{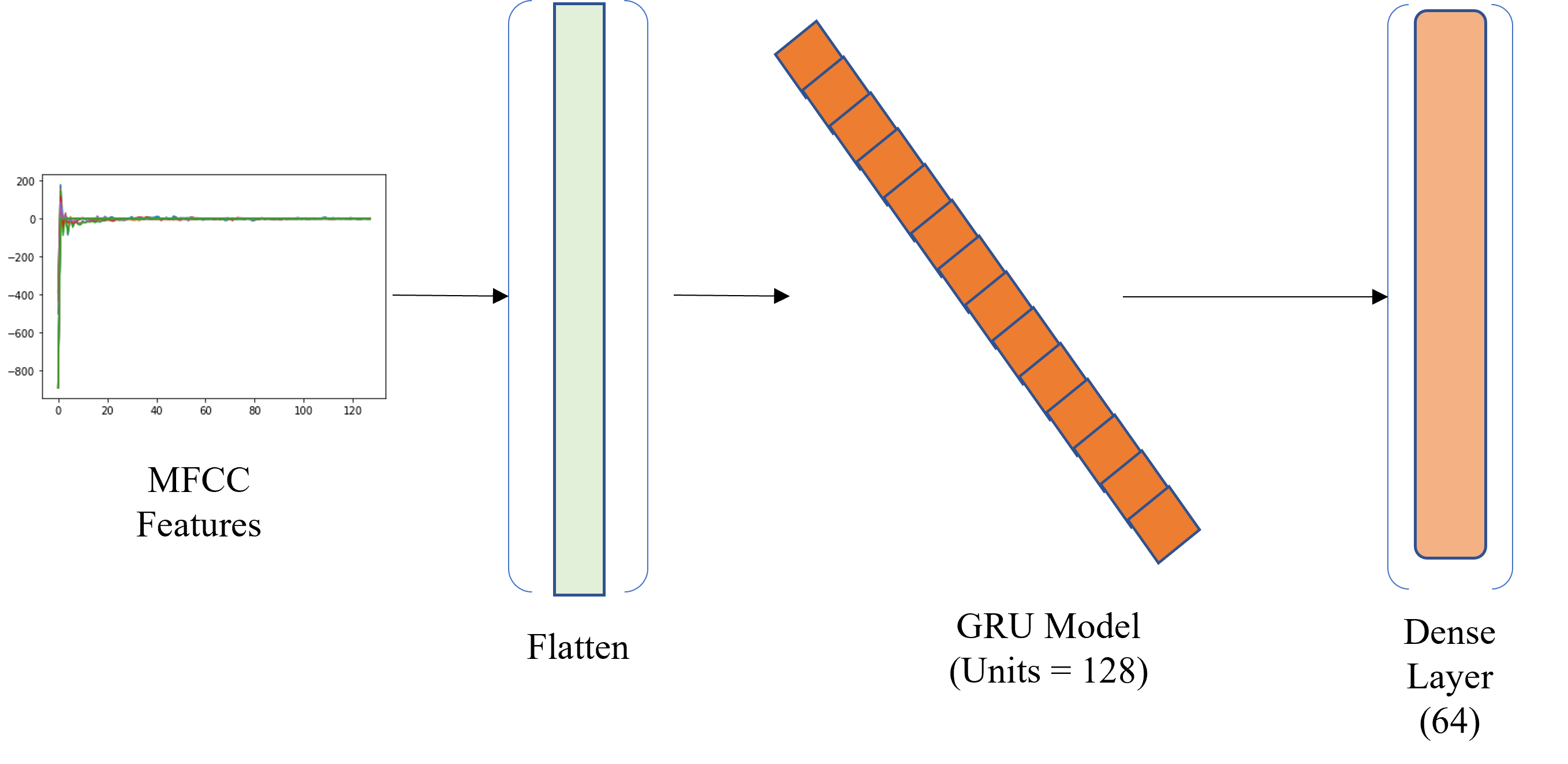}
    \caption{The detailed architecture of the recurrent stream}
    \label{gru}
\end{figure}

The MFCC features are first flattened before passing on to the recurrent network. For the recurrent network we use the GRU model \cite{cho2014learning}. Other RNN models such as LSTM and Transformers were tried in place of the GRU model but the network performed best using a GRU model. Here we used a GRU model of 128 units to learn the sequential dependencies of the MFCC features. The features generated from the gru model are then passed to a fully connected layer of 64 length. Both the stream uses a fully connected layer of 64 length to convert their features vectors to an equal length that is later used in the attention model. The recurrent network is presented in Figure \ref{gru}. 

\subsection{Attention Network}
\begin{figure}[h]
    \centering
    \includegraphics[width=\linewidth]{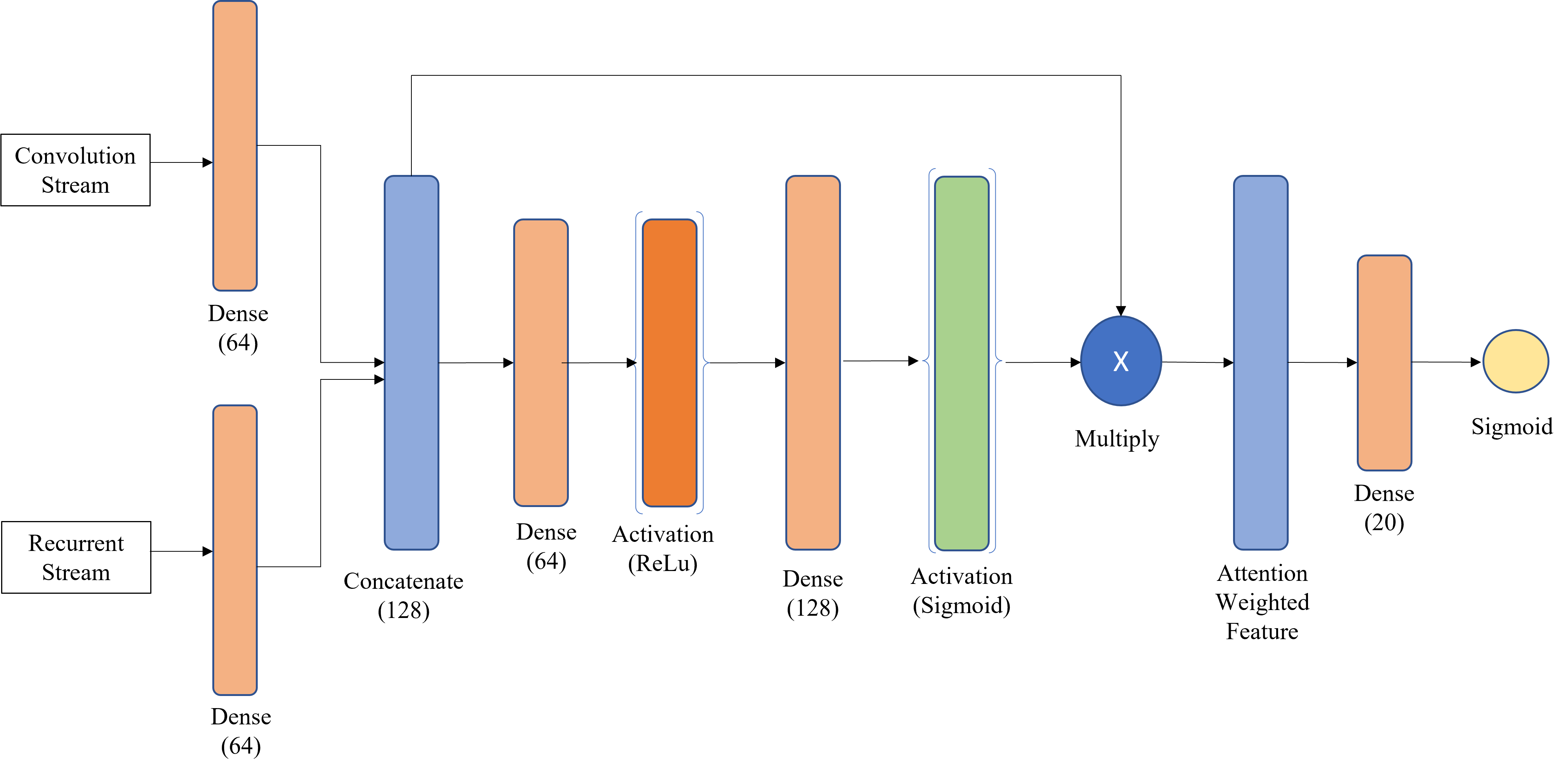}
    \caption{Network architecture of the attention module}
    \label{attention}
\end{figure}

The features from the convolutional stream and the recurrent stream are merged together using the proposed novel attention network. The equal sized feature vectors from both the streams are first concatenated. This concatenated feature vector of 128 length is then passed through a fully connected layer of 64 length and relu activation. By doing this we downsample the features vector and it is then again upsampled to a length of 128 using another fully connected layer which has a sigmoid activation function. Since this layer uses sigmoid activation, it produces a probabilistic feature space that assigns a value close to one to the features that are more valuable in the classification task and a value close to zero to the features that has little impact to the classification task. This feature space is then multiplied with the concatenated feature vector of both the streams. As a result, the concatenated feature vector now uses attention to strengthen the features that are more valuable for this task. It is then passed through two dense layers to make the final prediction. The attention network is presented in Figure \ref{attention}

\section{Experimentation and Results}
\subsection{Dataset}

The dataset used in this work is taken from the PhysioNet/Computing in Cardiology (CinC) Challenge 2016 that includes nine heart sound databases collected by several independent research teams around the world. \cite{liu2016open} The dataset consists of 3153 heart sound recordings from a total of 764 patients. It includes 2488 recordings collected from healthy subjects and 665 recordings from patients with various heart diseases such as coronary artery disease and heart valve defects. The length of the recordings varies from several seconds to several minutes. In this study 90\% of the data is used as the train set, and the rest of the 10\% data is used as the test set. 

\subsection{Experimental Setup}

The proposed network is trained for 50 epochs using a batch size of 128. During training, a balanced batch sampler is used to ensure that the same number of samples are taken from each class for every batch. Adam optimizer \cite{kingma2014adam} with a learning rate of 1e-3 and a binary crossentropy loss function is used to train the model.

\subsection{Results}

The performance of the proposed scheme is evaluated both on individual audio cycles and on a patient basis. During training, all the audio cycles extracted from a normal labeled patient were labeled as normal, and vice versa for the abnormal class. 

Four random train-test folds were generated from the dataset and the performance of the model was evaluated on all four folds. The results of the test set data of Fold-1 are presented in Table \ref{res:audio}. In this method, the sensitivity, specificity, accuracy, and MACC are found to be quite satisfactory, with values of \textbf{76.74\%}, \textbf{92.88\%}, \textbf{84.16\%}, and \textbf{84.81\%}, respectively.

\begin{table}[h]
\centering
\caption{Results of the Audio Cycles of both class on the test set data of Fold-1}
\label{res:audio}
\begin{tabular}{|c|c|c|c|}
\hline
Class    & Precision & Recall & F1-Score \\ \hline
Abnormal & 92.67\%    & 76.74\% & 83.96\%   \\ \hline
Normal   & 77.28\%    & 92.88\% & 84.37\%   \\ \hline
Average  & 84.98\%    & 84.81\% & 84.16\%   \\ \hline
\end{tabular}
\end{table}

But ultimately the sysem has to give predictions on a patient level. For this, the results of all the audio cycles of a patient are averaged and if the majority of the cycles are predicted as Normal then the patient is also predicted to be normal and vice versa for the abnormal class. This method further increases the performance of the proposed system, as can be observed from Table \ref{res:patient}. For the test set of Fold-1, this method achieves a very high sensitivity, specificity, accuracy, and MACC of \textbf{83.63\%}, \textbf{96.5\%}, \textbf{90.06\%}, and \textbf{90.07\%}, respectively. 

\begin{table}[h]
\centering
\caption{Results of Patient-Level of both class on the test set data of Fold-1}
\label{res:patient}
\begin{tabular}{|c|c|c|c|}
\hline
Class    & Precision & Recall & F1-Score \\ \hline
Abnormal & 95.97\%    & 83.62\% & 89.37\%   \\ \hline
Normal   & 85.49\%    & 96.49\% & 90.66\%   \\ \hline
Average  & 90.73\%    & 90.06\% & 90.02\%   \\ \hline
\end{tabular}
\end{table}

The patient-level results for all four folds are presented in Table \ref{res:fold}. The average of these values is determined to present a more general and robust valuation of the proposed system. On average, the model has accuracy, sensitivity, specificity, and MACC values of \textbf{87.11\%}, \textbf{82.41\%}, \textbf{91.8\%}, and \textbf{87.12\%}, respectively. 

\begin{table}[h]
\centering
\caption{Results on Patient-Level for all Four Folds}
\label{res:fold}
\begin{tabular}{|c|c|c|c|c|}
\hline
Fold No. & Accuracy & Sensitivity & Specificity & MACC   \\ \hline
Fold-1   & 90.06\%   & 83.63\%      & 96.5\%       & 90.07\% \\ \hline
Fold-2   & 81.48\%   & 83.44\%      & 79.45\%      & 81.45\% \\ \hline
Fold-3   & 88.89\%   & 81.29\%      & 96.5\%       & 88.89\% \\ \hline
Fold-4   & 88.02\%   & 81.29\%      & 94.74\%      & 88.02\% \\ \hline
\textbf{Average}  & \textbf{87.11\%}   & \textbf{82.41\%}      & \textbf{91.8\%}       & \textbf{87.12\%} \\ \hline
\end{tabular}
\end{table}

To demonstrate the effectiveness of the proposed network architecture, an ablation study was carried out on the system. Therefore, the different parts of the network were trained and tested on the Fold-1 dataset, and cycle level results were produced. It can be observed from Table \ref{res:ablation} that using the convolution stream alone resulted in an accuracy of 81.81\%, whereas using only the recurrent stream with raw data achieves an accuracy of only 57.71\%. When MFCC features were given as input to the recurrent stream as opposed to the raw data, the accuracy increased by 39.98\% to a value of 80.78\%. When both streams were combined, the accuracy improved by 3.79\% and became 83.84\%. Finally, the proposed method that combined the two streams with a novel attention module resulted in the best accuracy of \textbf{84.16\%}.

\begin{table}[h]
\centering
\caption{The ablation study result of the proposed method}
\label{res:ablation}
\begin{tabular}{|c|c|c|c|c|}
\hline
Method                                                                           & Accuracy & Sensitivity & Specificity & MACC    \\ \hline
\begin{tabular}[c]{@{}c@{}}Convolution \\ Stream\end{tabular}                    & 81.81\%  & 69.27\%     & 96.53\%     & 82.9\%  \\ \hline
\begin{tabular}[c]{@{}c@{}}Recurrent Stream \\ with Raw Data\end{tabular}        & 57.71\%  & 26.01\%     & 94.92\%     & 60.47\% \\ \hline
\begin{tabular}[c]{@{}c@{}}Recurrent Stream \\ with MFCC Feature\end{tabular}    & 80.78\%  & 70.19\%     & 93.21\%     & 81.7\%  \\ \hline
\begin{tabular}[c]{@{}c@{}}Dual Stream Network \\ without Attention\end{tabular} & 83.84\%  & 75.76\%     & 93.32\%     & 84.54\% \\ \hline
\textbf{Proposed Method}                                                                  & \textbf{84.16\%}  & \textbf{76.74\%}     & \textbf{92.88\%}     & \textbf{84.81\%} \\ \hline
\end{tabular}
\end{table}

In order to evaluate the performance of the proposed method, it was compared with other methods found in the literature that used the same PhysioNet 2016 dataset. \cite{liu2016open} The comparison result is reported in Table \ref{res:comp}. From the table, it can be concluded that the proposed scheme outperforms all the other methods in terms of accuracy and specificity, and gives a promising result in terms of sensitivity and $F_{1}$ score.

\begin{table}[h]
\centering
\caption{Comparison of the proposed method with the existing methods in the literature}
\label{res:comp}
\begin{tabular}{|c|c|c|c|c|}
\hline
Method                    & Accuracy & Sensitivity & Specificity & MACC   \\ \hline
Fan Li et al.\cite{li2020classification}         & -        & 87\%         & 86.6\%       & 86.8\%  \\ \hline
Ahmed et al.\cite{humayun2020towards}          & 80.39\%   & 87.68\%      & 71.23\%      & 79.46\% \\ \hline
Yi He et al.\cite{he2021research}          & -        & 96.4\%       & 78.1\%       & 87.3\%  \\ \hline
Baris Bozkur et al.\cite{bozkurt2018study} & 81.5\%    & 84.5\%       & 78.5\%       & 81.5\%  \\ \hline
\textbf{Proposed Method}           & \textbf{87.11\%}   & \textbf{82.41\%}      & \textbf{91.8\%}       & \textbf{87.12\%} \\ \hline
\end{tabular}
\end{table}

\section{Discussion}
From the result section it can be observed that the proposed system gives a satisfactory performance and it has a comparable result with the state of the art methods. The ablation study of the proposed system further confirms the effectiveness of the different modules that were introduced in this work. One of the main challenges of this dataset was the noisy signal that makes it difficult to distinguish between the normal and abnormal class. Even though the proposed system manages to learn from these noisy signals, in order to further improve the performance of the model the properties of the noisy signals have to be analyzed and learnable network modules have to be introduced that can mitigate the effect of the noise in the signal. 

\section{Conclusion}
An automatic detection system of heart abnormality from the heart sound signals promises to be a early screening tool that can have great impact in reducing heart disease related mortality. This research was conducted with the aim of developing a robust and generalized deep learning based system that can detect abnormality in the heart sound signal with high accuracy. It is shown that the proposed system utilizes the raw heart sound signals and its MFCC feature in a dual stream network and manages to effectively merge them using the novel attention network. With further improvement it is expected that the proposed system can one day be deployed in real life scenario and have an impact on people's life. 
\bibliographystyle{IEEEtran}
\bibliography{refs.bib}
\end{document}